\documentclass[11pt]{article}

\usepackage[active]{srcltx}
\usepackage{amsfonts}

\def\be{\begin{equation}}
\def\ee{\end{equation}}
\def\bea{\begin{eqnarray}}
\def\eea{\end{eqnarray}}
\def\noi{\noindent}
\def\nn{\nonumber}

\begin{document}

\begin{center}
{\Large {\bf  Contact Symmetries in Non-Linear Mechanics: a preliminary step to (Non-Canonical) Quantization}}
\end{center}

\bigskip
\bigskip

\centerline{V. Aldaya$^{1}$, J. Guerrero$^{2,1}$, F.F. L\'opez-Ruiz$^{3,1}$ and F. Coss\'{\i}o$^{1}$}

\bigskip
\centerline{\it$^1$ Instituto de Astrof\'\i sica de Andaluc\'\i a (IAA-CSIC),}
\centerline{\it Apartado Postal 3004, 18080 Granada, Spain}
\centerline{\it $^2$ Departamento de Matem\'atica Aplicada, Facultad de Inform\'atica,  Universidad de  Murcia,}
\centerline{\it Campus de Espinardo,  30100 Murcia, Spain}
\centerline{\it $^3$ Departamento de  F\'{\i}sica Aplicada, Universidad de C\'adiz,}
\centerline{\it Campus de Puerto Real, E-11510 Puerto Real, C\'adiz, Spain}
\bigskip

\centerline{valdaya@iaa.es $\quad$ juguerre@um.es $\quad$ paco.lopezruiz@uca.es $\quad$ fcossiop@gmail.com}

\bigskip

\begin{center}

{\bf Abstract}
\end{center}

\small
\setlength{\baselineskip}{12pt}

\begin{list}{}{\setlength{\leftmargin}{3pc}\setlength{\rightmargin}{3pc}}
\item
In this paper we exploit the use of symmetries of a physical system so as to characterize algebraically
the corresponding solution manifold by means of Noether invariants. This constitutes a necessary preliminary step towards
the correct quantization in non-linear cases, where the success of Canonical Quantization is not guaranteed in general.
To achieve this task ``point symmetries'' of the Lagrangian are generally not enough, and the notion of
contact transformations is in order: the solution manifold can not be in general parametrized by means of Noether
invariants associated with basic point symmetries. The use of the contact structure given by the Poincar\'e-Cartan
form permits the definition of the symplectic form on the solution manifold, through some sort of Hamilton-Jacobi
transformation. It also provides the required basic symmetries, realized as Hamiltonian vector fields associated
with global functions on the solution manifold (thus constituting an inverse of the Noether Theorem), lifted back
to the evolution space through the inverse of this Hamilton-Jacobi mapping. In this framework, solutions and symmetries,
as a whole, are somehow identified and this correspondence is also kept at a perturbative
level. We present  non-trivial examples of this interplay between symmetries and solutions pointing out the usefulness
of this mechanism in approaching the corresponding quantization. In particular, we achieve the proper generalization of the Heisenberg-Weyl algebra
for the non-linear particle sigma model in $S^3$ within this framework, and notice
that a subset of the classical symmetries corresponding with this quantizing algebra
(those generalizing boosts) are necessarily of non-point character.
\end{list}

\normalsize

\noi {\it Keywords}: Symmetry; non-linear systems; non-point symmetries; Cartan formalism; Hamilton-Jacobi;
      inverse Noether theorem; $S^3$ sigma model.

\setlength{\baselineskip}{14pt}

\section{Introduction: Basic symmetries}

Symmetries have played an important role in Physics and their systematized study can be traced back to pioneers works by
Lie, B\"acklund, Noether, etc. in Classical Mechanics (see for instance \cite{Noether1918,Ibragimov1985,Stephani1989}) or Weyl, Wigner,
Bargmann, etc. since the beginning of Quantum Mechanics \cite{Weyl,Wigner,Waerden,Bargmann}. They contributed
the classification of solutions and/or the generation of new ones. But there is a very relevant question in dealing with
non-linear system which can be addressed by making an exhaustive use of symmetries; that is, the global characterization of
(the set of solutions of) a given physical system or a class of equivalent systems both at the classical or at the quantum levels. This global
characterization will manifest itself as a preponderant task in finding a consistent (let us say correct) quantization of
soundly non-linear system. Many fundamental non-linear systems are related with gravity and Non-Linear Sigma Models; maybe the difficulty
in quantizing Gravity is more related to its non-linearity than its {\it gravity-ness}.

In this paper we shall deal with some aspects of symmetries that
are not, in practice, familiar to symmetry-workers and these
aspects rely on the very power of this instrument. We refer to the
ability of symmetries in collecting themselves into a Lie group
containing a co-adjoint orbit that mimics the symplectic Solution
Manifold (SM) of the symmetric system (see, in particular,
\cite{Souriau}, where the name ``space of motions" is used). This
property, which is relevant in defining globally the (Poisson)
structure of a classical system, states the base of a consistent
quantization; after all, a quantization of a physical system is
supposed to be a unitary and irreducible representation, much in
the sense of Lie groups, of a (basic) subalgebra of the Poisson
algebra defined in its SM. The actual way of working out this
representation, in particular the representation space interpreted
as the Hilbert space of quantum states, does basically not matter.
Canonical Quantization does obey this perspective provided that
the basic symmetry group can be identified with the
Heisenberg-Weyl group.

As regards the relevant symmetry  for Canonical Quantization, that
is, a Heisenberg-Weyl subalgebra of the Poisson algebra, it must
be stressed that half the classical functions used to define the
(basic) canonical Poisson brackets (initial position of the
particle) are (Noether invariants) associated with transformations
of the ``boost'' type and these transformations do not leave, in
general, invariant (not even semi-invariant, i.e. up to a total
derivative) the Lagrangian driving the physical problem. We
usually face transformations in the ``Evolution Manifold'', EM for
short (we shall name so the space of time, ordinary co-ordinates
and velocities or momenta where the system actually evolves;
the name ``evolution space" is also used in the literature), which are
not derived from transformations of just the space-time.
Traditionally this kind of transformations are referred to as
``contact transformation'', as opposed to ``point
transformations'', and are rarely considered in Physics. Contact
symmetries are those which leave semi-invariant (that is, up to a
total differential) the Poincar\'e-Cartan  form $\Theta_{PC}=
\frac{\partial {\cal L}}{\partial\dot{q}}(dq-\dot{q}dt)+{\cal
L}dt$, that is, a generalization of the action integrand which
reduces to ${\cal L}dt$ when the Lagrangian ${\cal L}$ is regular,
since in this case $\dot{q}$ is necessarily $\frac{dq}{dt}$ on the
solutions. The price to be paied for considering  a basic Poisson
subalgebra among classical functions associated with non-point
symmetries is that the corresponding quantum observables,
according to Canonical Quantization, would be, in general, of
non-local type, that is, they would contain arbitrary powers of the
standard momentum. Possible ordering problems are definitely
avoided if we adopt a non-canonical scheme for quantization like
Geometric Quantization \cite{Souriau} or, even better, Group
Approach to Quantization \cite{GAQ} (see also \cite{Vallareport}
and references there in) where the basic quantum operators are
directly generators of a Lie group irrespective of the actual
analytical expressions of the corresponding Noether invariants.

The global characterization of the SM in  group-algebraic terms (as the co-adjoint orbit of a dynamical group in the sense of Ref. \cite{Souriau})
brings the extra bonus that it allows us to relate strongly
physical systems which evolve in a different way. We can in fact
relate systems sharing the same SM. In this situation the
evolution of those systems will also share isomorphic symmetries.
In this way, we shall generalize to non-linear systems the
algorithm associated with the Arnold Transformation \cite{Arnoldo}
(which has been recently extended to the quantum case
\cite{ArnoldQ}) intended to relate the properties of the free
particle with those of a physical problem driven by a linear
second-order differential equation with arbitrary, time-dependent
coefficients. The sharing of the solution manifolds can then be
seen as an algorithm for finding symmetries of a given problem by
importing them from a simpler one (the free particle if the SM
were ``so flat''). It should also be remarked, however, that in
order to share symmetries we have to know the whole set of
solutions of both systems. This is quite clear because the
knowledge of the SM for any system requires its solutions.
Fortunately, all the considerations above can be kept at a
perturbative level, so that finding perturbative solutions and
perturbative symmetries can be achieved with the beat.

The present paper is organized as follows. In Sec. II we review
the general, though basic aspects of Classical Mechanics, lending
special attention to those objects which are going to play a
relevant role in the algebraic structure of symmetries, that is,
Poincar\'e-Cartan form, contact symmetries, Hamilton-Jacobi
transformation and the passing to the SM, Liouville and symplectic
forms in the SM and Hamiltonian symmetries. In Sec. III we analyze
the sharing of symmetries and the ``lifting'' of symmetries from
the SM to the EM. In this respect, we study the general procedure
to achieve this task perturbatively by means of the Magnus series.
In Sec. IV some paradigmatic examples are considered in different
subsections including the anharmonic oscillator and the
$S^3$-sigma model, where the proper Heisenberg-Weyl-like group is found.
Finally in Sec. V we briefly outline the
parallel treatment of the present symmetry interplay at the
quantum level.

\section{The structure of Classical Mechanics}

In that which follows we shall pay special attention to the basic concepts and algorithms, trying to avoid technicalities as
much as possible. For more precise and complete reports we refer the reader to
\cite{AbrahamMarsden,ArnoldClassicalMechanics,Goldstein,Landau,RNC,Sarlet} and references therein.
We shall think primarily  of Classical Mechanics
in the tangent space to a manifold $\Sigma$, the cofiguration space, added with time, $EM \equiv \mathbb{R}\times T(\Sigma)$, with coordinates $(t,q,\dot{q})$,
although much work can easily be translated to Classical Field Theory by using
coordinates $(x^\mu,\phi(x),\phi_\mu(x))$ associated with fields $\phi$, and their formal derivatives $\phi_\mu(x)$, on
a space-time $M=\mathbb{R}\times S$ (and the integration volume $dt$ on $\mathbb{R}$ with $\omega$ on $M$)
\footnote{Technically,
$EM$ is the 1-jet bundle of the bundle $\mathbb R \times \Sigma \rightarrow \mathbb R$, to be replaced by the 1-jet bundle of a bundle
$E \rightarrow \mathbb R \times S$, where $E$ is the fiber bundle whose sections are the fields and $S$ is some Cauchy surface.}
; see for instance \cite{Higgs}.

The Hamilton Principle (HP) establishes that the solutions of a variational problem characterized by the Lagrangian ${\cal L}$ are those
curves on $\Sigma$ for which the action functional
\[ {\cal S}[q(t)]=\int {\cal L}(t, q,\dot{q}=\frac{dq}{dt})dt \]
has an extreme. If the Lagrangian is regular, this happens when $q(t)$ satisfies the Euler-Lagrange equations
\[ \frac{d}{dt}\frac{\partial {\cal L}}{\partial\dot{q}}-\frac{\partial {\cal L}}{\partial q}=0 \,,\]
and it must be noticed that the Lagrangian, as a function on $\mathbb{R}\times T(\Sigma)$, depends on $t$, $q$, and $\dot{q}$,
where $\dot{q}$ parameterizes possible values of time derivatives of curves but does not correspond to any specific one.
Symmetries in the context of the HP are transformations of $\mathbb{R}\times\Sigma$ which leave invariant the action functional ${\cal S}$, or
just semi-invariant (up to a total derivative) the integrand ${\cal L}dt$. In infinitesimal terms, the
transformation in $\mathbb{R}\times\Sigma$ is generated by a vector field
\[ Y=Y^t\frac{\!\!\partial}{\partial t}+Y^q\frac{\!\!\partial}{\partial q} \;\; \:\:(Y^t\approx \delta t, Y^q\approx \delta q) \]
and the Lie derivative of ${\cal L}$ with respect to the natural prolongation of $Y$ from $\mathbb{R}\times\Sigma$ to $\mathbb{R}\times T(\Sigma)$, that is, the
``1-jet-extension'' $\bar{Y}$ (in the language of jet-bundles; see, for instance \cite{Hermann,Aldaya19}), is a total derivative:
\[
L_{\bar{Y}}{\cal L}=\frac{df_Y}{dt}  \,.
\]
The 1-jet-extension $\bar{Y}$ is $Y$ added with the variation of $\dot{q}$, that is,
$\bar{Y}^{\dot{q}}\frac{\!\!\partial}{\partial\dot{q}}$, where
\[  \bar{Y}^{\dot{q}}= \frac{d Y^q}{d t}-\frac{d Y^t}{d t}\dot{q} \]
which means that $\delta\dot{q}$ behaves as $\frac{\!\!d}{dt}\delta q$.

The Modified Hamilton Principle (MHP) generalizes the HP in that the variations on $q$ and $\dot{q}$ (or $p$) are independent \cite{Aldaya19}
(in basic textbooks of Mechanics, like \cite{Goldstein,Landau}, this variational principle was considered, but also in more
elaborate texts devoted to Gravity, like \cite{MisnerThorneWheeler}, to account for Palatini's formalism).  Under the MHP, physical
trajectories are critical points of a modified action ${\cal S}^1$ which is a functional of curves on  $T(\Sigma)$:
\[ {\cal S}^1[q(t),\dot{q}(t)]=\int_{[q(t),\dot{q}(t)]} \Theta_{PC} \]
where the Poincar\'e-Cartan(-Hilbert) form, $\Theta_{PC}$ is defined as
\[ \Theta_{PC}=\frac{\partial {\cal L}}{\partial \dot{q}}(dq-\dot{q}dt)+{\cal L}dt \,, \]
so that it reduces to ${\cal L}dt$ if we restrict the arguments of ${\cal S}^1$ to curves that are 1-jet-extension of curves on $\Sigma$,
that is, curves for which $\dot{q}=\frac{dq}{dt}$.

The variations of ${\cal S}^1$ are written in terms of the Lie derivative of the integrand with respect to an arbitrary vector field $X^1$
on $\mathbb{R}\times T(\Sigma)$, non-necessarily of the form $\bar{X}$:
\[\delta{\cal S}^1(X^1)|_{[q(t),\dot{q}(t)]}=\int L_{X^1}\Theta_{PC}=0\,, \]
and lead to the equations of motion (the symbol $i_X$ on a differential form stands for the {\it interior product}, that is,
the form applied to the vector field)
\begin{equation}
 i_{X^1}d\Theta_{PC}|_{[q(t),\dot{q}(t)]}=0 \:\:\: \forall X^1 \: \hbox{on} \: \mathbb{R}\times T(\Sigma)\,, \label{equation}
\end{equation}
since the second term in the (Cartan) decomposition of the Lie derivative $i_{X^1}d+di_{X^1}$ under integration gives
rise to a total differential.  When the Lagrangian is regular, the equation associated with $\delta\dot{q}$ says that
$\dot{q}$ on the trajectory is actually $\frac{dq}{dt}$, and that associated with $\delta q$ reproduces the standard
Euler-Lagrange equation for the curve $q(t)$.  In this regularity situation, in which the Legendre transformation is invertible,
the Poincar\'e-Cartan form acquires the more traditional expression (canonical or Darboux-like)
\[ \Theta_{PC}= pdq-Hdt,\: \: H=\dot{q}p-{\cal L},\:\: p\equiv\frac{\partial {\cal L}}{\partial\dot{q}} \]

The introduction of the Poincar\'e-Cartan form, even for the case of regular Lagrangians, proves to be specially useful.
On the one hand, in characterizing symmetries and proving the

\medskip

\noindent \underline{\it Noether Theorem}: If a vector field $Y^1$ on $\mathbb{R}\times T(\Sigma)$ is a symmetry of the variational problem,
that is, if $L_{Y^1}\Theta_{PC}=df_{Y^1}$ for some function $f_{Y^1}$ associated with $Y^1$, then the quantity $F_{Y^1}\equiv i_{Y^1}\Theta_{PC}-f_{Y^1}$
is a constant along the solutions.

\noindent The proof is evident after having written the equations of motion as in (\ref{equation}).

\medskip

This general definition of ({\it contact}) symmetries obviously contains the standard ({\it point}) symmetries, that is,
when $Y^1$ is the prolongation $\bar{Y}$ of a vector field $Y$ on $\mathbb{R}\times\Sigma$.

On the other hand, the Poincar\'e-Cartan form defines the {\it Poisson Structure} on the set of solutions of the physical
problem defined by ${\cal L}$. Given two Noether Invariants $F_{Y^1}$ and $F_{Z^1}$ associated with two symmetries $Y^1$ and $Z^1$,
respectively, their Poisson bracket is defined as the Noether Invariant associated with the Lie bracket of the respective symmetries:
\begin{equation}
\{F_{Y^1},F_{Z^1}\}\equiv i_{[Y^1,Z^1]}\Theta_{PC}-f_{[Y^1,Z^1]}\equiv F_{[Y^1,Z^1]}  \,. \label{Poisson}
\end{equation}
This definition generalizes (and reproduce in many cases) the standard prescription
\[ \{p,q\}=1\,. \]

After the (intrinsic) definition above (\ref{Poisson}), we wish to remark that the Poisson bracket is defined solely between
constants of motion; this ``restriction'' is required in order to make the pairing $\{\;,\;\}$ invertible. In fact, the inverse of this pairing
$\{\;,\;\}^{-1}$ is, by definition, the {\it Symplectic form} $\Omega$ on the set of solutions, i.e., the {\it Solution Manifold} $SM$ .
To be precise, given two generators of symmetries $Y^1$ and $Z^1$, we define
\begin{equation}
 \Omega(Y^1,Z^1)\equiv \{F_{Y^1},F_{Z^1}\} = d\Theta_{PC}(Y^1,Z^1)\,.\label{Simplectica}
\end{equation}
It must be remarked that the $2$-form $d\Theta_{PC}$ itself does not define a symplectic structure
on $EM$ because it
has a non-trivial kernel generated by the equations of motion. Notice that the trajectories of the variational
problem (\ref{equation}) can also be seen, directly,  as the curves generated by the vector fields in the kernel
of $d\Theta_{PC}$, which may contain additional vector fields (other than the one associated with the time evolution properly)
if the Lagrangian is not regular.

The vector fields on $EM\equiv \mathbb R \times T(\Sigma)$ that are symmetries define vector fields on $SM$. For this reason we shall keep the notation $Y^1$, $Z^1$, etc. in both cases.

Let us now assume, for the sake of simplicity, that the Lagrangian is regular, so that the Poincar\'e-Cartan form
can be written as
\[ \Theta_{PC}=pdq-Hdt \]
and the kernel of $d\Theta_{PC}$ is generated by just the  evolution in time. Adopting for this generator the expression
\begin{equation}
\check{X}_H\equiv\frac{\!\!\partial}{\partial t}+ X^q\frac{\!\!\partial}{\partial q}+X^p\frac{\!\!\partial}{\partial p}\,,
\end{equation}
that is, the component in $t$, $X^t$, equals one (a different choice would only lead to a time redefinition), we obtain
\be
\check{X}_H = \frac{\!\!\partial}{\partial t}+\frac{\partial H}{\partial p}\frac{\!\!\partial}{\partial q}-
\frac{\partial H}{\partial q}\frac{\!\!\partial}{\partial p} \,,
\label{Hamiltonian}
\ee
reproducing the standard Hamilton equations of motion $\frac{d t}{ds}=1$, $\frac{d q}{ds}= \frac{\partial H}{\partial p}$,
$\frac{d p}{ds}=- \frac{\partial H}{\partial q}$.

We proceed to realize a ``change of co-ordinates'' which makes apparent the kernel of $d\Theta_{PC}$, allowing for an explicit
writing of co-ordinates in the {\it Symplectic Solution Manifold} $SM$. In other words, we take the quotient in the Evolution Manifold
by the equations of motion. This process is guaranteed by the Frobenius Lemma \cite{Godbillon}, which ensures the existence of co-ordinates $(s,Q,P)$ in which
the vector field (\ref{Hamiltonian})  is written as
\[   \check{X}_H=\frac{\!\!\partial}{\partial s} \,.\]
In fact, we perform the change of variables
\begin{eqnarray}
t &=& \varphi^0(Q,P,s)=s\nn\\
q &=& \varphi(Q,P,s)\nn\\
p &=& \varphi^*(Q,P,s)\,,\label{HJ}
\end{eqnarray}
where $\varphi$ and $\varphi^*$ are the solutions of the equations of motion, with initial constants $Q,P$,
\begin{eqnarray}
 \frac{d\varphi}{ds}&=&X^q(q,p,t)\nn\\
 \frac{d\varphi^*}{ds}&=&X^p(q,p,t)\nn\\
 \frac{dt}{ds}&=&X^t(q,p,t)=1\,,
\end{eqnarray}
and computing $\frac{\!\!\partial}{\partial s}$ with the Jacobian of (\ref{HJ}) we obtain ($\frac{dQ}{dt}=0,\,\frac{dP}{dt}=0$):
\begin{eqnarray}
 \frac{dq}{dt}&=&\frac{\partial q}{\partial Q}\frac{dQ}{dt}+\frac{\partial q}{\partial P}\frac{dP}{dt}+
     \frac{\partial q}{\partial s}\frac{ds}{dt}=\frac{\partial q}{\partial s}=X^q\nn\\
 \frac{dp}{dt}&=&\frac{\partial p}{\partial Q}\frac{dQ}{dt}+\frac{\partial p}{\partial P}\frac{dP}{dt}+
     \frac{\partial p}{\partial s}\frac{ds}{dt}=\frac{\partial p}{\partial s}=X^p\nn\\
 \frac{\!\!\partial}{\partial s}&=&\frac{\partial t}{\partial s}\frac{\!\!\partial }{\partial t}+
     \frac{\partial q}{\partial s}\frac{\!\!\partial }{\partial q}+
 \frac{\partial p}{\partial s}\frac{\!\!\partial }{\partial p}=\frac{\!\!\partial}{\partial t}+
     X^q\frac{\!\!\partial}{\partial q}+X^p\frac{\!\!\partial}{\partial p} \,.\nn
\end{eqnarray}

The change of variables (\ref{HJ}) can be named the {\it
Hamilton-Jacobi transformation} (it is usually referred to as the
Canonical Transformation that takes the co-ordinates $q,p$ to
constant co-ordinates $Q,P$) and is an invertible mapping in the
Evolution Manifold, $EM=\mathbb{R}\times T(\Sigma)$, which turns the new time variable into a mere
spectator as far as the Solution Manifold is concerned. In fact,
the forms $\Theta_{PC}$ and $d\Theta_{PC}$ are written in the new
co-ordinates as though time were absent,
\bea
\Theta_{PC}&=&pdq-Hdt=PdQ+d(S(t,q,p)-PQ)\,,\nn\\
d\Theta_{PC}&=&dP\wedge  dQ \,,
\eea
except for a total derivative of a function, $S-PQ$, where $S$ is the Hamilton Principal function \cite{Goldstein}.
They define on $SM$ the symplectic form $\Omega=dP\wedge  dQ$, for which
the 1-form $\Lambda=PdQ$, called the
Liouville form, behaves as a symplectic potential for $\Omega= d\Lambda$. The explicit form of
the transformation (\ref{HJ}) is not unique because we could
choose different constants instead of $Q,P$, thus allowing for non-Canonical Transformations generalizing the usual concept;
they are only
intended to be Noether invariants uniquely determined by $Q,P$.
The reader may have in mind the simplest example of the free particle,
where $Q=q-\frac{p}{m}t$ is the Noether invariant associated with
the Galilean boost, and $P=p$ is the Noether invariant associated with
translations. In this example, the equation (\ref{HJ}) reads $t=s,\, q=Q+\frac{P}{m}s,\,p=P$.

Once again we mention that the Hamilton-Jacobi transformation is a
change of co-ordinates in the EM that suggests the quotient
\[ (\mathbb{R}\times T(\Sigma),d\Theta)/\check{X}_H\approx (SM,\Omega) \]
of the original Lagrangian system by the equations of motion, and that it is in $(SM,\Omega)$
where (Analytical) Mechanics is properly realized
and where Quantum Mechanics starts from.
We also employ the name Hamilton-Jacobi for the projection associated with the Hamilton-Jacobi transformation.

Let us end this section by pointing out that the symmetries of our
physical system must be characterized by vector fields on $SM$
leaving invariant the symplectic form $\Omega$ and being globally
Hamiltonian, that is, the symmetries are generated by vector
fields $X_f$, associated with a function $f$ actually defined on
$SM$, satisfying:
\be
  i_{X_f}\Omega=-df \;\;\Rightarrow L_{X_f}\Omega=0\,,\label{hamiltonianvector}
\ee
where $f$ is some function of $Q,P$. In particular, and given the ``canonical'' (Darboux)
co-ordinates $Q,P$ on $SM$, the local expression of $X_f$ adopt the standard form
\[ X_f= \frac{\partial f}{\partial P}\frac{\!\!\partial}{\partial Q}-
\frac{\partial f}{\partial Q}\frac{\!\!\partial}{\partial P}\,,
 \]
and the natural local (``canonical'') basic symmetries are
those generated by the vector fields
\[ \frac{\!\!\partial}{\partial Q}, -\frac{\!\!\partial}{\partial
P}\]
with Hamiltonian functions $P$ and $Q$ respectively. These functions close a Heisenberg-Weyl
algebra but, in general, it is not globally defined.

For the sake of precision, we should comment something about the
globality of the Hamilton-Jacobi transformation. This will be
ensured whenever the vector field generating the equations of
motion is in turn a generator of a Lie group that characterizes
completely the system. We shall always pursue this situation.

\section{Symmetries from the Solution Manifold and the Sharing of Symmetries}

The Liouville and the symplectic forms on $SM$ have been written
in a co-ordinate system $(Q,P)$ that provide them with their
canonical expression.  This  is always permitted, at least locally, as a
consequence of Darboux's theorem. In general, and depending on the
topological structure of $SM$, it could be more convenient to
choose (part of the) co-ordinates as (Noether invariants) directly
associated with transformations globally defined on the EM (and $SM$). In particular, if
$SM$ can be recovered as a co-adjoint orbit of a Lie group, the co-ordinates $(Q,P)$ should
then be replaced by (the local expressions of) Noether invariants associated with a (``basic'') subgroup generating
the orbit (some group generalizing the Heisenberg-Weyl group \cite{Bargmann,Levy-Leblond,GAQ}). Even more, new $Q'(Q,P)$ and
$P'(Q,P)$ might be chosen so as to close a finite-dimensional Poisson subalgebra with the Hamiltonian
(see for instance \cite{pasteles}).

As mentioned above, any (globally will be assumed hereafter) Hamiltonian vector field on $SM$ is a proper symmetry
of the physical system and any function $f$ on $SM$ generates a symmetry through the corresponding
Hamiltonian vector field $X_f$ as given by (\ref{hamiltonianvector}). Now, by inverting the
Hamilton-Jacobi transformation we arrive at an
\medskip

\noi \underline{\it Inverse Noether Theorem}, which states that given a constant of the motion, that is, any function $f$ on $SM$,
the corresponding Hamiltonian vector field $X_f$, written in the original co-ordinates $(t, q, p)$, i.e.
transformed by the inverse of the Hamilton-Jacobi transformation (\ref{HJ}), is a (contact in general)
symmetry of the Poincar\'e-Cartan form $\Theta_{PC}$. This can be thought of as a ``lifting'' of a symmetry from $SM$ to $EM$.

Of special significance are the symmetry $X_H$, associated with an autonomous Hamiltonian $H$ on the EM, $\mathbb{R}\times T(\Sigma)$,
which is written in $SM$ by keeping its functional dependence, and the symmetries associated with the
Hamiltonian functions $Q, P$. In computational terms, we only have to work out the Jacobian of (\ref{HJ}),
\bea
\frac{\!\!\partial}{\partial s}&=&\frac{\partial t}{\partial s}\frac{\!\!\partial }{\partial t}+
     \frac{\partial q}{\partial s}\frac{\!\!\partial }{\partial q}+
 \frac{\partial p}{\partial s}\frac{\!\!\partial }{\partial p}\nn\\
\frac{\!\!\partial}{\partial Q}&=&\frac{\partial t}{\partial Q}\frac{\!\!\partial }{\partial t}+
     \frac{\partial q}{\partial Q}\frac{\!\!\partial }{\partial q}+
 \frac{\partial p}{\partial Q}\frac{\!\!\partial }{\partial p}\nn\\
\frac{\!\!\partial}{\partial P}&=&\frac{\partial t}{\partial P}\frac{\!\!\partial }{\partial t}+
     \frac{\partial q}{\partial P}\frac{\!\!\partial }{\partial q}+
 \frac{\partial p}{\partial P}\frac{\!\!\partial }{\partial p}\,,\nn
\eea
and, if as already supposed $t=s$, just the matrix elements $\frac{\partial q}{\partial Q},
\frac{\partial p}{\partial Q}, \frac{\partial q}{\partial P}, \frac{\partial p}{\partial P}$.

It must be noted that the lifted symmetries in $EM$ are in what is sometime called ``evolutionary form'' \cite{Evolutionary}, i.e,
they do not contain terms in $\frac{\!\!\partial}{\partial t}$. However, the possibility exists of adding a term of the form
$\chi(Q,P,s)\frac{\!\!\partial}{\partial s}$ to any vector field on $SM$, with a proper choice of the function $\chi$,
in order to achieve on the EM a more traditional, though equivalent form of the symmetry.
This can be
illustrated with the Hamiltonian $X_H$, itself, which can be ``lifted'' to the EM in the
traditional  form as corresponding with the time translation:
\[ \frac{\!\!\partial}{\partial s}-X_H=\frac{\!\!\partial}{\partial t} \,.\]
This can also be used to make explicit the ``geometrical'' character of some generators, i.e. their $1$-jet prolongation from $\mathbb{R}\times\Sigma$
to the whole $EM$.

In coming explicitly from the EM  down to  $SM$ the time variable can be factored out and the physical content
of the dynamical system relies entirely on the Hamiltonian $H$, in much the same way the Lagrangian describes
the physical content of the problem formulated in the EM. From the point of view of the
Poisson algebra on $SM$, however, many different Hamiltonian functions could be chosen to recreate distinct time evolutions.
In other words, from the SM manifold, in abstract, we may construct many evolution manifolds, each one
associated with the inverse Hamilton-Jacobi transformation corresponding in turn with different
Hamiltonians. Those EM's are, by definition, sharing the same SM and, therefore, the same symmetries
(written in different ways though), preserving the same Lie algebra structure. The situation can be depicted
by means of the following diagram involving two different physical systems $(EM_1, {\cal L}_1)$, $(EM_2, {\cal L}_2)$ constructed
from the same SM by the inverse Hamilton-Jacobi transformation associated with two different Hamiltonians $H_1$ and $H_2$:

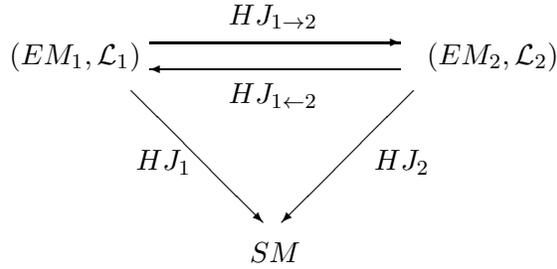
\begin{picture}(350,135)(0,0)
\put(72,100){$(EM_1, {\cal L}_1)$}
\put(125,108){\vector(1,0){95}}
\put(220,98){\vector(-1,0){95}}
\put(230,100){$(EM_2, {\cal L}_2)$}
\put(155,115){$HJ_{1\rightarrow 2}$}
\put(155,85){$HJ_{1\leftarrow 2}$}
\put(118,90){\vector(1,-1){50}}
\put(225,90){\vector(-1,-1){50}}
\put(163,25){$SM$}
\put(120,60){$HJ_1$}
\put(210,60){$HJ_2$}
\end{picture}

It constitutes an algorithm to export  symmetries from the
physical system, say $(EM_1, {\cal L}_1)$, to the other system
$(EM_2, {\cal L}_2)$, and of particular relevance is the case
where the system $1$ is simpler (free particle, harmonic
oscillator, for instance) and the system $2$ is described by a
(regular) Lagrangian, the Hamiltonian of which can be written in
the $SM$ as that of the free system, $H_0$, plus an ``interaction
term'' $H_I$, usually depending only on $Q$.

As described above, the characterization of a given physical system is achieved by means of its solutions and/or basic symmetries,
symplectic structure and/or Poisson brackets, etc. and all this can be realized clearly either on $(EM,\Theta_{PC})$, restricted
to solutions and symmetries, or on  $(SM,\Omega)$ after having taken explicitly the quotient by the evolution $\check{X}_H$. In
the first case we might speak of a description ``\`a la Schr\"odinger'', by importing the language from Quantum Mechanics, whereas
in the second, we should then refer to this realization as a description  ``\`a la Heisenberg''. The actual recreation of time
evolution from the SM, or that which is the same, the explicit construction of the inverse HJ transformation is obviously a
non-trivial task and requires, in general, a perturbative scheme to compute the exponential of the Hamiltonian $X_H$, acting
on functions on $SM$ as a derivation on this manifold or, equivalently, by the Poisson action with the function $H$. Given
an arbitrary function $f$ on $SM$ the evolutive version of $f$ on the EM can be seen as a solution of the first-order differential equation
\[ \frac{\!\!d}{ds}f=\{f,H\}\equiv -X_Hf \]
the solution of which can be formally written as a series
expansion
\be F(t,q,p)=U(t)F_0=e^{\Omega(t)}F_0\,\,, F_0\equiv
F(0,q,p)\equiv f(Q,P)\,\,, \label{Magnus} \ee
where
\bea
   \Omega^{[0]}\equiv0\,\,,\Omega (t) &=& \lim_{n\to\infty} \Omega^{[n]}(t)\\
\Omega^{[n]} (t) &=& \sum_{k=0}^{\infty}\frac{B_k}{k!}\int^{t}_{0}
dt_1
   {\rm ad}^k_{\Omega^{[n-1]}(t_1)}
       \bigl(-X_H\bigr)
\label{eq:Omegan}\,, \eea
$B_k$ are the Bernoulli numbers and
\begin{equation}
{\rm ad}^0_{A} (\cdot)\equiv  \cdot\,, \qquad {\rm ad}^1_{ A} (
\cdot)\equiv [A, \cdot]\,, \qquad {\rm ad}^k_{ A}( \cdot)\equiv
[A,\,{\rm ad}^{k-1}_{ A}(\cdot) ]\,,\label{Magnus2}
\end{equation}
where $A$ is an arbitrary vector field and $\cdot$ stands for the
argument to which the $ad_A$ is applied (-$X_H$ in
(\ref{eq:Omegan})).

The expansion above is known as the Magnus Series
\cite{Valencianos} and differs from the more traditional
Dyson-like perturbative series in that the former keeps
``unitarity'' even at finite orders (at the classical level it
should be called ``simplecticity'' in the sense that the
integration volume in $SM$ is preserved); see \cite{Manolo} for
more details. For a time-independent Hamiltonian, as corresponds
to a function defined on $SM$, we arrive at a rather simple
formula for evolved (inverse Hamilton-Jacobi transformed)
functions:
\begin{equation}
  F(t)  =
    \sum_{k=0}^{\infty} \frac{1}{k!}(-t)^k
     {\rm ad}^k_{X_H} \bigl(F_0 \bigr)\,.
\end{equation}
This evolutive development, when applied to the basic functions
$Q,P$, constitutes the explicit form of the inverse
Hamilton-Jacobi transformation, $HJ(H)^{-1}$, from $SM$ to the EM,
for a general Hamiltonian $H$.

The perturbative computation (\ref{Magnus}) entails some practical inconvenience when the Hamiltonian $H$ is seen
as the sum of a simpler Hamiltonian $H_0$, to be referred to as non-perturbed Hamiltonian, and an interaction term
to be denoted $H_I$, both originally defined on $SM$ as functions of $Q,P$. It is due to the fact that the power
series in the time evolution parameter parallels the power series in the coupling constant $\lambda$ to which $H_I$
were proportional, that is, $H$ is not a homogeneous function of the coupling constant. To take advantage of the
facility of fully integrating the time evolution corresponding to the non-perturbed Hamiltonian, we may resort to
the trick of performing the inverse Hamilton-Jacobi transformation in two steps: we go from $SM$, parameterized by $(Q,P)$,
to an Evolution Manifold $EM_0$, parameterized by $(t, q_0, p_0)$, through the inverse Hamilton-Jacobi
transformation given by $H_0$, and then consider $EM_0$ as though it where the
Solution Manifold from which to evolve with some sort of interaction Hamiltonian $H_I(t)\equiv JH(H_0)^{-1}H_I$,
which now depends on time and, therefore, the Magnus series must be computed according to the general formula (\ref{Magnus}).
This situation can be illustrated with a particular case of the picture above, that is,

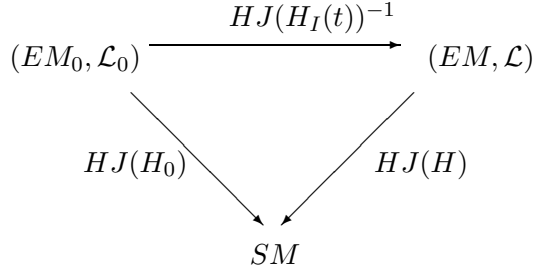
\begin{picture}(350,135)(0,0)
\put(72,100){$(EM_0, {\cal L}_0)$}
\put(125,108){\vector(1,0){95}}
\put(230,100){$(EM, {\cal L})$}
\put(155,115){$HJ(H_I(t))^{-1}$}
\put(118,90){\vector(1,-1){50}}
\put(225,90){\vector(-1,-1){50}}
\put(163,25){$SM$}
\put(100,60){$HJ(H_0)$}
\put(210,60){$HJ(H)$}
\end{picture}

\noi and the description of the physical system addressed by $H$ in co-ordinates $(t,q_0,p_0)$ may be referred to as
the  ``Interaction Picture'' (this method is similar to the one discussed in \cite{Souriau}, due to Lagrange, and known as method of ``variation of the constants'').

\section{Examples of Basic Symmetries in Non-Linear Systems}

In this Section we present some examples explicitly. Although part of their symmetries are perhaps well known,
the characterization of the symplectic solution manifold requires additional non-trivial (and not so well-known) symmetries.

\subsection{Examples with trivial topology}

In this subsection we consider a couple of examples bearing non-trivial symmetries, in the sense that they
are not all point symmetries, but evolving in an EM with trivial topology so that the choice of co-ordinates does
not require of special care.

\subsubsection{Free relativistic particle}

Here we desire to point out the relevance of non-point symmetries
by demonstrating that even in the example of a free relativistic
particle  the true position variables are associated with
this kind of symmetries, as opposed to the boost symmetries which
are of the ``ordinary'' type. We consider just the one-dimensional case for the sake
of simplicity.

The Lagrangian for the free relativistic particle of mass $m$ is given by ($c$ is the speed of light)
\[ {\cal L}=-mc^2\sqrt{1-\frac{\dot{q}^2}{c^2}}\]
and, it being regular, we shall assume $\dot{q}=\frac{dq}{dt}$ on
trajectories. The momentum is $p=\frac{\partial {\cal
L}}{\partial\dot{q}}=m\dot{q}/\sqrt{1-\frac{\dot{q}^2}{c^2}}$.
From it we derive the Poincar\'e-Cartan form
\be
\Theta_{PC}= \frac{m\dot{q}}{\sqrt{1-\frac{\dot{q}^2}{c^2}}}dq-\frac{mc}{\sqrt{1-\frac{\dot{q}^2}{c^2}}}dt= pdq-p_0cdt\,,
\ee
where $cp_0\equiv c\sqrt{p^2+m^2c^2}=mc^2/\sqrt{1-\frac{\dot{q}^2}{c^2}}$ is the (the energy) Hamiltonian $H$.

The equations of motion are derived as usual, as the trajectories of the kernel of $d\Theta_{PC}$, that is, the integral curves of
\[ \check{X}_H= \frac{\!\!\partial}{\partial t}+\frac{pc}{p_0}\frac{\!\!\partial}{\partial q}\]
with solutions
\bea
t&=&\varphi^0(Q,P,s)=s\nn\\
q&=&\varphi(Q,P,s)=\frac{c\,P}{P_0}s+Q\\
p&=&\varphi^*(Q,P,s)=P\,,\nn
\eea
which constitutes the HJ transformation ($P_0=p_0$). Rewriting the Poincar\'e-Cartan form in the new co-ordinates $(s,Q,P)$ we obtain
the canonical expression up to a total differential:
\[ \Theta_{PC}= pdq-cp_0dt=PdQ-d(\frac{m^2c^3s}{P_0})\,.\]
The basic symmetries, with the corresponding Noether invariants, are then:
\bea
X_P\equiv\frac{\!\!\partial}{\partial Q}&=&\frac{\!\!\partial}{\partial q}\;\;\Rightarrow \;\; F_P=P=p\nn\\
X_Q\equiv-\frac{\!\!\partial}{\partial P}&=&-\frac{\!\!\partial}{\partial p}-\frac{m^2c^3t}{p_0^3}\frac{\!\!\partial}{\partial q}=
   -\frac{1}{mc}(1-\frac{\dot{q}^2}{c^2})^{\frac{3}{2}}[ct\frac{\!\!\partial}{\partial q}+
    c\frac{\!\!\partial}{\partial \dot{q}}]\;\;\Rightarrow\;\; F_Q=Q=q-\frac{p}{p_0}ct\nn
\eea

We observe that the symmetry $\frac{\!\!\partial}{\partial P}$ is actually a non-point symmetry associated with the true position
(initial position). In fact, the Lagrangian is not even semi-invariant under $X_Q$:
\[L_{X_Q}{\cal L}=(1-\frac{\dot{q}^2}{c^2})\dot{q}\,,\]
which is not a total derivative. However, the Poincar\'e-Cartan form is left semi-invariant:
\[ L_{X_Q}\Theta_{PC}=d(q-\frac{\dot{q}^2}{c^2}\dot{q}t)\,.\]
Note in passing that this non-point symmetry is usually not considered in the literature.

Much better known is the related  symmetry, the one corresponding
to the relativistic boost, which is found by lifting from $SM$ the
Hamiltonian vector field
\[ X_K=\frac{P_0}{mc}\frac{\!\!\partial}{\partial P}-\frac{PQ}{mcP_0}\frac{\!\!\partial}{\partial Q}\,,\]
and that, although non-point symmetry at first sight, can be added
with the extra term $\chi\partial/\partial s=
(Q+P/P_0s)\partial/\partial s$, as commented in the previous
section, to read on the EM as
\[ \frac{q}{c}\frac{\!\!\partial}{\partial t}+ct\frac{\!\!\partial}{\partial q}+c(1-\frac{\dot{q}^2}{c^2})\frac{\!\!\partial}{\partial \dot{q}}\;\;\Rightarrow\;\;F_K\equiv K=\frac{p_0}{mc}q-\frac{p}{mc}ct\]
where $K=QP_0/mc$ is the ordinary boost constant.

It is clear that the Noether constants $P,K,H$ close a Poisson subalgebra, the Poincar\'e Lie algebra, whereas $P,Q,H$ does not, although
in passing to the quantum theory the operators associated with these last functions are devoid of unitarity
problems (see \cite{Newton-Wigner}, for instance).

We could also go further in selecting a finite-dimensional Poisson subalgebra in $SM$ by redefining both the momentun $P$ and
the position $Q$ in the form:
\be
 \Pi\equiv\frac{2mcP}{\sqrt{2mc(P_0+mc)}}\;,\;\;\;\mathbb{X}\equiv\frac{2P_0Q}{\sqrt{2mc(P_0+mc)}} \label{posicion}
\ee
so that the energy $H$, with the rest mass subtracted (this quantity permits a proper $c\rightarrow\infty$ limit) is written
as $H-mc^2=\frac{\Pi^2}{2m}$. Since the Poisson bracket of the new variables is also $\{\Pi,\,\mathbb{X}\}=1$,
the set of functions $(\mathbb{X},\,\Pi,\,H,\, 1)$ closes, surprisingly, a (centrally extended) Galilean algebra
\bea
\{\Pi,\,\mathbb{X}\}&=&1\nn\\
\{H,\,\mathbb{X}\}&=&\Pi\nn\\
\{H,\,\Pi\}&=&0\nn
\eea
that can be realized, in passing to the EM, as a (non-point, though) symmetry of the relativistic particle. In particular, the position variable
in (\ref{posicion}) is essentially the Hamiltonian function which gives rise to the mean position operator in the minimal, canonical
representation of the Poincar\'e group for spin zero. From the point of view of Canonical Quantization, it behaves as a non-local  operator
and, in fact, it was originally obtained, for arbitrary spin, along with the conserved mean spin operator, through the non-local
Foldy-Wouthuysen transformation \cite{Foldy-Wouthuysen} (see also the book \cite{Fonda}).


\subsubsection{Anharmonic oscillator: sharing symmetries with the harmonic one}


The present example of an harmonic oscillator of mass $m$ and
angular frequency $\omega$, perturbed by an interaction of the
form $\lambda\frac{q^4}{4}$, is a simple case of an integrable
system (in the Liouville sense), whose solutions can be expressed
in terms of Jacobi Elliptic functions, that nevertheless can be
treated perturbatively  as a useful illustration of the
``Interaction Picture'' algorithm described above. (In fact, the
present treatment would also be applied as such to a more general
perturbation of the form $\lambda\frac{q^n}{n}$, for arbitrary
integer $n$; we restrict $n=4$, just for concreteness). We shall
try to implement the basic symmetries of the unperturbed harmonic
oscillator, that is, the Newton-Hook group with (Poisson) Lie
algebra,
\bea
\{P,\,Q\}&=&1\nn\\
\{H_0,\,Q\}&=&P/m\label{Newton-HookPoisson}\\
\{H_0,\,P\}&=&\omega^2 Q\nn
\eea
where $H_0=P^2/2m+1/2m\omega^2Q^2$ is the corresponding Hamiltonian written in the Solution Manifold, as a symmetry
of the perturbed system described by the complete Hamiltonian $H=p^2/2m+1/2m\omega^2q^2+\lambda\frac{q^4}{4}$. Since
$H$ is conserved, it is intended to be written as $H=P^2/2m+1/2m\omega^2Q^2+\lambda\frac{Q^4}{4}\equiv H_0+H_I(Q)$ in the SM, which is
shared with that of the unperturbed system.

Let us denote the variables in the EM$_0$ of the unperturbed system by $(t,\,q_0,\,p_0)$, with Hamiltonian $H_0=p_0^2/2m+1/2m\omega^2q_0^2$, and
write explicitly the Hamilton Jacobi transformation $HJ(H_0)$ associated with the unperturbed Hamiltonian,
\bea
t&=&\tau\nn\\
q_0&=&Q\cos(\omega t)+ \frac{P}{m\omega}\sin(\omega t)\nn\\
p_0&=&P\cos(\omega t)-m\omega Q \sin(\omega t)\nn
\eea
as well as the vector fields (lifted expressions of the Hamiltonian vector fields associated with
$H_0$,$Q$, $P$), generating the Newton-Hook (point) symmetry:
\bea
X_t&=&\frac{\!\!\partial}{\partial t}\nn\\
X_{q_0}&\equiv& X_Q = \cos(\omega t)\frac{\!\!\partial}{\partial q_0}-m\omega \sin(\omega t)\frac{\!\!\partial}{\partial p_0}\label{NewtonHookLibre}\\
X_{p_0}&\equiv& X_P= \frac{\sin(\omega t)}{m\omega}\frac{\!\!\partial}{\partial q_0}+\cos(\omega t)\frac{\!\!\partial}{\partial p_0}\nn
\eea

We now proceed to construct order by order the transformation $HJ(H_I(t))^{-1}$ from the Evolution Manifold of the unperturbed harmonic oscillator to
the Evolution Manifold of the perturbed one taking into account the explicit time dependence of $H_I(t)$ when written in terms of
the variables in EM$_0$:
\[H_I\equiv\frac{\lambda}{4}Q^4=\frac{\lambda}{4}(q_0\cos(\omega t)-\frac{p_0}{m\omega}\sin(\omega t))^4\,.\]
Making explicit use of equations (\ref{Magnus}-\ref{Magnus2}), applied to the functions $q_0,\,p_0$, up to the order $n$, and expressing back the result in
terms of the $SM$ variables we obtain the $n$th-order approximation of the solutions of the anharmonic oscillator. In particular, for $n=1$, we arrive at
the result:
\bea
q(t)&=& q_0 +
\frac{\lambda}{32 m^4 \omega^5}\left[ -9 m\omega p_0^2 q_0  +12\omega p_0^3 t  -5 m^3\omega^3 q_0^3 +12 m^2\omega^3 p_0 q_0^2 t  -8 p_0^3 \sin (2  \omega t) \right. \nn\\
& & +\omega\cos (2  \omega t) \left(4 m^3 q_0^3 \omega^2+12 m p_0^2 q_0  \right) + \omega\cos (4  \omega t) \left(m^3 q_0^3 \omega^2-3 m p_0^2 q_0  \right) \nn \\
& & \left. +\sin (4  \omega t) \left(p_0^3-3 m^2\omega^2 p_0 q_0^2 \right) \right] + O(\lambda^2)\nn \\
&=& Q \cos (\omega t )+\frac{P }{m \omega }\sin ( \omega t ) \nn\\
& &+  \frac{\lambda}{32 m^4 \omega ^5}\left[
\omega\cos ( \omega t) \left(-m^3 \omega^2  Q^3 +12 m^2 \omega^2 P Q^2 t +3 m P^2 Q  +12 P^3 t  \right) \nn \right.\\
& &+\sin ( \omega t) \left(-12 m^3 Q^3 t \omega ^4-21 m^2 P Q^2 \omega ^2-12 m P^2 Q t \omega ^2-9 P^3\right) \nn\\
& & \left.+\omega\cos (3  \omega t) \left(m^3 \omega^2 Q^3-3 m P^2 Q  \right)+\sin (3  \omega t) \left(3 m^2\omega^2 P Q^2 -P^3\right)
  \right] + O(\lambda^2) \label{Solutions} \\
p(t)&=&p_0 -
\frac{\lambda}{32 m^3 \omega ^4}\left[12 m^3\omega^4 q_0^3 t -15 m^2\omega^2 p_0 q_0^2 +12 m\omega^2 p_0^2 q_0 t -3 p_0^3  +8 m^3\omega^3 q_0^3  \sin (2  \omega t) \right.\nn \\
& & +\cos (2  \omega t ) \left(12 m^2\omega^2 p_0 q_0^2 +4 p_0^3\right) +\cos (4  \omega t) \left(3 m^2\omega^2 p_0 q_0^2 -p_0^3\right)\nn \\
& & \left. +\omega\sin (4  \omega t) \left(m^3\omega^2 q_0^3 -3 m p_0^2 q_0  \right) \right] + O(\lambda^2) \nn \\
& = & P \cos ( \omega t )-m\omega Q   \sin ( \omega t ) \nn\\
& & -\frac{\lambda}{32 m^3 \omega^4}\left[
 \omega\sin ( \omega t) \left(11 m^3\omega^2 Q^3 +12 m^2\omega^2 P Q^2 t +15 m P^2 Q  +12 P^3 t \right)  \right.\nn\\
& & +\cos ( \omega t) \left(12 m^3\omega^4 Q^3 t +9 m^2 \omega^2 P Q^2+12 m\omega^2 P^2 Q t -3 P^3\right) \nn \\
& & +\omega\sin (3  \omega t) \left(3 m^3\omega^2 Q^3 -9 m P^2 Q  \right)
\left. +\cos (3  \omega t ) \left(3 P^3-9 m^2\omega^2 P Q^2 \right)
\right]  + O(\lambda^2) \nn
\,.
\eea

%
The  Jacobian of the transformation in terms of the unperturbed variables $(q_0,p_0)$
brings the  vector fields (\ref{NewtonHookLibre}) to the  perturbed Evolution
Manifold,
\bea
X_t&=&\frac{\!\!\partial}{\partial t}
-\lambda\frac{\sin (t \omega ) }{m^3 \omega^3}
\left(q  \cos (t \omega )-\frac{p}{m\omega} \sin (t \omega )\right){}^3\frac{\!\!\partial}{\partial q}\nn\\
& & -\lambda\frac{\cos (t \omega ) }{m^2 \omega^2}
\left(q   \cos (t \omega )-\frac{p}{m\omega} \sin (t \omega )\right){}^3\frac{\!\!\partial}{\partial p} \nn\\
X_{q_0}&=& \cos(\omega t)\frac{\!\!\partial}{\partial q}-m\omega \sin(\omega t)\frac{\!\!\partial}{\partial p}
+\frac{3\lambda}{32 m^3 \omega ^4}
\Bigl(
\cos ( \omega t) \left(-3 m^2\omega^2 q^2 +8 m \omega^2 p q t+3 p^2\right)\Bigr.\nn\\
& & +\omega\sin ( \omega t) \left(-4 m^2\omega^2 q^2 t +10 m p q  -12 p^2 t  \right)
+\cos (3  \omega t) \left(3 m^2\omega^2 q^2 -3 p^2\right) \nn \\
& &  \Bigl. -6 m\omega p q   \sin (3  \omega t)
\Bigr)\frac{\!\!\partial}{\partial q}
-\frac{3 \lambda}{32 m^2 \omega^3}\Bigl(
 \omega\cos ( \omega t) \left(12 m^2\omega^2 q^2 t -6 m  p q  +4 p^2 t  \right)\Bigr.\nn\\
& &+\sin ( \omega t) \left(11 m^2 q^2 \omega ^2-8 m p q t \omega ^2+5 p^2\right)
+6 m\omega p q   \cos (3  \omega t) \nn \\
& & \Bigl. +\sin (3  \omega t) \left(3 m^2\omega^2 q^2 -3 p^2\right)
\Bigr)\frac{\!\!\partial}{\partial p}  + O(\lambda^2) \label{NewtonHook}\\
X_{p_0}&=&\frac{\sin(\omega t)}{m\omega}\frac{\!\!\partial}{\partial q}
+\cos(\omega t)\frac{\!\!\partial}{\partial p}
 +\frac{3 \lambda}{32 m^4 \omega^5}
\Bigl(
\omega\cos ( \omega t) \left(4 m^2\omega^2 q^2 t -2 m p q  +12 p^2 t  \right)\Bigr.\nn\\
& & +\sin ( \omega t) \left(-7 m^2\omega^2 q^2 +8 m\omega^2 p q t -9 p^2\right)
+2 m\omega p q   \cos (3  \omega t) \nn\\
& &\Bigl. +\sin (3  \omega t) \left(m^2 \omega^2 q^2-p^2\right)
\Bigr)\frac{\!\!\partial}{\partial q}
+\frac{3\lambda }{32 m^3 \omega^4}
\Bigl(
\cos ( \omega t) \left(-m^2\omega^2 q^2 -8 m\omega^2 p q t +p^2\right)\Bigr.\nn\\
& & +\omega\sin (\omega t ) \left(-12 m^2\omega^2 q^2 t +14 m p q  -4 p^2 t  \right)
+\cos (3  \omega t) \left(m^2 q^2 \omega ^2-p^2\right) \nn \\
& & \Bigl.-2 m\omega p q   \sin (3  \omega t)\Bigr)\frac{\!\!\partial}{\partial p} + O(\lambda^2)\nn \,,
\eea
which can be confirmed as non-point symmetries of the
Poincar\'e-Cartan form of the anharmonic oscillator except for
$O(\lambda^{2})$, closing the Newton-Hook Lie algebra
(\ref{Newton-HookPoisson}) up to $O(\lambda^{2})$. Repeating the procedure for higher values of $n$ we would obtain
a power series in $\lambda$ for the vector fields $X_t, X_{q_0}, X_{p_0}$ which would be non-point, exact symmetries
of the Poincar\'e-Cartan form, and which would close the Newton-Hook algebra exactly.

\subsection{Example with non-trivial topology: Particle sigma-model}

A class of paradigmatic examples of non-linear physical systems
corresponds to the motion of a particle  on a group manifold $G$,
which plays the role of configuration space. The evolution
manifold is then $EM=\mathbb{R}\times T(G)$. $G$ is usually a
semi-simple (mostly compact) Lie group endowed with the
(pseudo)-Riemannian metric given by the right-invariant (or
left-invariant) canonical $1$-form $\theta^R$ (or $\theta^L$), as
follows
\be
g_{ij}=\theta^{(r)}_i\theta^{(s)}_jk_{rs}\,,\label{gij}
\ee
where $i,j,r,s$ run from $1$ to the dimension of $G$, parameterized by local canonical 
co-ordinates at the identity, say $\{\epsilon^i\}$, $k_{rs}$ is a invertible bilinear 
symmetric form on the Lie algebra, and $\theta^{(i)}_s$ is the inverse matrix of the 
right-(or left-)invariant generators
$X_{(j)}=X_{(j)}^s\frac{\!\!\partial}{\partial \epsilon^s}$. In the more usual case 
of semi-simple Lie groups, $k_{rs}$ is taken to be the Killing form, so 
that $k_{rs}= C_{rm}^n C_{sn}^m$, where $C_{im}^n$ are
the structure constants. The Lagrangian driving the motion on $G$ is given by
\be
{\cal L}= \frac{1}{2} g_{ij}\dot{\epsilon}^i\dot{\epsilon}^j\,. \label{lagrangianoSigma}
\ee

Sigma models were firstly introduced in low-energy treatment of strong interactions (see for instance \cite{Gell-MannLevy})
which were mediated by fields
$(\pi^+,\pi^0,\pi^-,\sigma)$ restricted by a quadratic relation among the pion field $\vec{\pi}$
and the auxiliary field $\sigma$ which gave name to those models, also called chiral models
because the $R$ and $L$ upstairs can be interchanged.
In fact, one can state immediately the invariance of (\ref{lagrangianoSigma})
under right-invariant vector fields (generators of the left
translation on $G$) as well as left-invariant vector fields (generators of the right translation),
which provide equivalent  Noether invariants (people refer to this chirality by saying that the
Lagrangian above is invariant under $G \times G$). Either symmetries generalize the ordinary
translations in flat space and constitute half of the required symmetries to characterize the
solution manifold. Some other symmetries playing the role of (generalized) boosts are then in order.

The difficulty in dealing with non-linear sigma models lies not
only on their non-linearity, clearly apparent by the dependence of
the metric $g$ on the coordinates $\epsilon$, but also on the
non-trivial topology of the configuration space $G$, determined by
its being semi-simple. The advantage, on the contrary, is the fact
that Lie groups contain canonical structures in terms of which the
physical quantities and the entire physical dynamics can be
expressed in an absolute way with independence of the choice of
local co-ordinates. In fact, the Lagrangian, Poincar\'e-Cartan form as
well as the Noether invariants associated with the generalized
translations along the non-flat geometry $G$ are neatly intrinsic.


\subsubsection{$S^3$ particle sigma-model}

The simplest example containing all the ingredients of Non-Linear
Sigma Models is that of a particle moving on the group $G=SU(2)$,
that is, with configuration space the three-dimensional sphere
$S^3$, and evolution manifold $\mathbb R \times T(S^3)$. We shall
parameterize the group manifold with co-ordinates $\{\epsilon^i\},
i=1,2,3$ determining the axis of a rotation $R(\vec{\epsilon}\,)$,
and the angle $\varphi$ through the relation
$|\vec{\epsilon}\,|=2\hbox{sin}\frac{\varphi}{2}$ with $0\le|\vec{\epsilon}|\le2$.
As mentioned
above we may disregard the fact that all the expressions are going
to be local provided that we keep ourselves working with canonical
structures defined on the Lie group.

Let us write the group law for $SO(3)$ ($\approx SU(2)$ locally) in the form
$\vec{\epsilon}\,''=f(\vec{\epsilon}\,',\vec{\epsilon}\,)$ from the
matrix multiplication
$R(\vec{\epsilon}\,'')=R(\vec{\epsilon}\,')R(\vec{\epsilon}\,)$:
\[ R(\vec{\epsilon}\,)^i_j=(1-\frac{\vec{\epsilon}\,^2}{2})\delta^i_j-\sqrt{1-
\frac{\vec{\epsilon}\,^2}{4}}\eta^i_{.jk}\epsilon^k+\frac{1}{2}\epsilon^i\epsilon^m
k_{jm}  \]
\be
\epsilon''^i=\sqrt{1-\frac{\vec{\epsilon}\,^2}{4}}\epsilon'^i+\sqrt{1-
\frac{\vec{\epsilon}\,'^2}{4}}\epsilon^i+\frac{1}{2}\eta^i_{.jk}\epsilon'^j\epsilon^k\,,
\ee
where $\eta^i_{.jk}$ is the completely anti-symmetric symbol in
three dimensions and $k_{jm}$ is the trivial Killing metric
$\delta_{jm}$, and derive the right-invariant vector fields as
well as the components of the right canonical $1$-form (the script
$R$ will be omitted after that):
\bea
X_{(i)}^R& = & X^k_{(i)}\frac{\!\!\partial}{\partial\epsilon^k}\equiv\left(\sqrt{1-\frac{\vec{\epsilon}\,^2}{4}}\delta^k_i+
    \frac{1}{2}\eta^k_{.im}\epsilon^m\right)\frac{\!\!\partial}{\partial\epsilon^k}\nn\\
{\theta^R}^{(i)}&=&\theta^{(i)}_jd\epsilon^j\equiv \left(\sqrt{1-\frac{\vec{\epsilon}\,^2}{4}}\delta^i_j+
     \frac{\epsilon^i\epsilon_j}{4\sqrt{1-\frac{\vec{\epsilon}\,^2}{4}}}-\frac{1}{2}\eta^i_{.jk}\epsilon^k\right)d\epsilon^j\,.
\eea

By making explicit the expression of the metric (\ref{gij}), and the (normalized) Killing form $k_{ij}=\delta_{ij}$, we
write the Lagrangian for the $S^3$-particle Sigma Model,
\be
{\cal
L}=\frac{1}{2}g_{ij}\dot{\epsilon}^i\dot{\epsilon}^j=\frac{1}{2}\left[\delta_{ij}+\frac{\epsilon_i\epsilon_j}{4(1-
    \frac{\vec{\epsilon}\,^2}{4})}\right]\dot{\epsilon}^i\dot{\epsilon}^j\,,\label{Lagrangian}
\ee
as well as the momenta, the Poincar\'e-Cartan form and the Hamiltonian:
\bea
p_i&=&\frac{\partial {\cal L}}{\partial\dot{\epsilon}^i}=g_{ij}\dot{\epsilon}^j\nn\\
\Theta_{PC}&=&g_{ij}\dot{\epsilon}^jd\epsilon^i-\frac{1}{2}g_{ij}\dot{\epsilon}^i\dot{\epsilon}^jdt=p_id\epsilon^i-Hdt\nn\\
H&=&{\cal L}=\frac{1}{2}(g^{-1})^{ij}p_ip_j=\frac{1}{2}\left(\delta^{ij}-\frac{\epsilon^i\epsilon^j}{4}\right)p_ip_j\,.\nn
\eea
From now on indices $i,j,k,...$ will be moved up and down with the trivial metric $k_{ij}$ and the use of the metric $g_{ij}$ will be
explicitly marked.

This Lagrangian (\ref{Lagrangian}) is known to be invariant under the (jet-prolongation of the right-invariant)
generators of the $SU(2)$ group
\be
\bar{X}_{(i)}=X_{(i)}^k \frac{\!\!\partial}{\partial\epsilon^k}+
     \frac{\partial X^k_{(i)}}{\partial\epsilon^s}\dot{\epsilon}^s \frac{\!\!\partial}{\partial\dot{\epsilon}^k}
\ee
with Noether invariants
\[ F^{(i)}=\theta^{(i)}_j\dot{\epsilon}^j\equiv\theta^i\,,\;\hbox{or}\;\;\theta_i\equiv X_{(i)}^kp_k\,. \]
These constants prove to be specially suited to parametrize the
flat piece of the SM as they are globally defined irrespective of
the group parameterization ($\vec{\epsilon}$ in this case). They
can replace the momenta $\{p_i\}$, according to their relations
above, although the price to be paid is the loss of canonical
Poisson bracket relationships. In fact, if the topology was
trivial, the basic Poisson bracket would have adopted the expression
$\{p_i,\epsilon^j\}=\delta^j_i$, but, in this case,  they will
acquire the non-canonical form (as associated with a globally
defined symplectic structure on the group)
\be
 \{\theta_i,\epsilon^j\}=X_{(i)}^j(\vec{\epsilon}\,) \label{chochetemalo}
\ee
although both brackets must be read from the SM. Thus we should
seek additional symmetries of the Poincar\'e-Cartan form providing
Noether invariants, in particular,
$\varepsilon^j\equiv\epsilon^j(t=0)$,  permitting  us to rewrite
(\ref{chochetemalo}) properly as a relationship between constants
of motion ($\vartheta_i\equiv\theta_i(0)=\theta_i$),
\be
\{\vartheta_i,\varepsilon^j\}=X_{(i)}^j(\vec{\varepsilon}\,)\,,
\label{chochetebueno} \ee
its r.h.s. being, as we are going to see, a linear combination o
the functions $\varepsilon^i$ and just one new function
$\rho(\vec{\varepsilon})$; they will close a seven-dimensional
Poisson subalgebra.

In fact, since we know the actual solutions of our dynamical system this task can be achieved by performing the corresponding HJ transformation,
which takes the Poincar\'e-Cartan form to $SM$, and by looking there for these extra symmetries, to be brought back to the EM. In fact,
the equations of motion,
\[ \frac{d\theta^i}{dt}=0\]
have exact solutions ($\omega\equiv\sqrt{2H}$)
\bea
\epsilon^i&\equiv&\epsilon^i(\vec{\varepsilon},\vec{\vartheta},s)=\varepsilon^i\hbox{cos}(\omega s)+\frac{\dot{\varepsilon}^i}{\omega} \hbox{sin}(\omega s)\nn\\
\theta^i&\equiv&\theta^i(\vec{\varepsilon},\vec{\vartheta},s)=\vartheta^i\label{soluciones}\\
(\dot{\epsilon}^i&\equiv&\dot{\epsilon}^i(\vec{\varepsilon},\vec{\dot{\varepsilon}},s)=\dot{\varepsilon}^i\hbox{cos}(\omega s)-
             \varepsilon^i\omega\hbox{sin}(\omega s))\nn\\
t&=&s\,,\nn
\eea
from which we read the HJ transformation. Note that the $SM$ is parameterized by $(\varepsilon^i, \vartheta_j)$.

Making use of formula (\ref{soluciones}), the Poincar\'e-Cartan form $\Theta_{PC}$ can be brought down to $SM$ turning into the Liouville
$1$-form $\Lambda$,
as well as its differential, reproducing the symplectic $2$-form, and the Hamiltonian function itself:
\bea
\Lambda&=&\vartheta_i\vartheta^{(i)}\nn\\
\Omega&=&d\vartheta_i\wedge\vartheta^{(i)}+\frac{1}{2}\eta^i_{.jk}\vartheta_i\vartheta^{(j)}\wedge\vartheta^{(k)}\label{symplectica}\\
H&=&\frac{1}{2}\vartheta_i\vartheta^i\,.\nn
\eea
In these expressions $\vartheta^i$ play the role of non-canonical momenta and $\vartheta^{(i)}\equiv\vartheta^{(i)}_jd\varepsilon^j$ the role of
{\it non-total differentials} of co-ordinates. However, if desired, we may rewrite $\Lambda$ and $\Omega$ in Darboux's form as
\[\Lambda=\pi_id\varepsilon^i,\;\;\Omega=d\pi_i\wedge d\varepsilon^i\,, \]
where $\pi$ is defined by $\vartheta_i\equiv X^k_i(\vec{\varepsilon})\pi_k$, although these expressions are valid only locally.

It is remarkable the fact that the expressions (\ref{symplectica}) prove to be quite general, valid for any semi-simple Lie group provided
that generic structure constants $C^i_{jk}$ substitute $\eta^i_{.jk}$.

Let us write the pairing, which associates functions on $SM$ with Hamiltonian vector fields, particularized to the basic functions
$\vartheta_i$ and $\varepsilon_j$:
\bea
i_{\mathbb{X}_{(i)}}\Omega&=&-d\vartheta_i\;\;\Rightarrow \;\; \mathbb{X}_{(i)}=X^k_{(i)}\frac{\!\!\partial}{\partial\varepsilon^k}-
    \frac{\partial X^k_{(i)}}{\partial\varepsilon^j}\pi_k\frac{\!\!\partial}{\partial\pi_j}\label{campossimetria}\\
i_{\mathbb{Y}_{(j)}}\Omega&=&-d\varepsilon_j\;\;\Rightarrow\;\; \mathbb{Y}_{(j)}=\frac{\!\!\partial}{\partial\pi_j}\,,
\eea
along with the function $\rho\equiv\sqrt{1-\frac{\vec{\varepsilon}\,^2}{4}}$:
\be
i_\mathbb{Z}\Omega=-d\rho\;\;\Rightarrow\;\; \mathbb{Z}=-\frac{\varepsilon_j}{\rho}\frac{\!\!\partial}{\partial\pi_j}
\ee

This set of vector fields close a $7$-dimensional Lie algebra, globally defined (as they correspond to complete Hamiltonian vector fields),
which constitutes the basic symmetry of our Physical system generalizing the Heisenberg-Weyl one,
which can only be locally defined.
The corresponding Poisson subalgebra closes, accordingly, as follows:
\bea
\{\varepsilon^i,\;\vartheta_j\}&=&\frac{1}{2}\eta^i_{.jk}\varepsilon^k+\delta ^i_j\rho\nn\\
\{\rho,\;\vartheta_k\}&=&-\frac{1}{4}\varepsilon_k\label{simetria}\\
\{\vartheta_i,\;\vartheta_j\}&=&\frac{1}{2}\eta^k_{.ij}\vartheta_k\,.\nn
\eea

It must be noticed that it is the Poisson subalgebra (\ref{simetria}) of the general Poisson algebra on $SM$
which intrinsically and minimally characterizes the classical system and, therefore, constitutes the starting point for a
proper (non-canonical) quantization of this typically non-linear system endowed with a non-trivial topology. The main virtue of the co-ordinates
$(\varepsilon^i,\,\vartheta_j)$, versus Darboux's ones, is precisely to make the topology of $SM$ manifest even locally. On the contrary, the
would-be ``canonical'' co-ordinates $(\varepsilon^i, \,\pi_j)$ prompt to a naive canonical quantization,  which would require
special attention in order to address unitarity. In a forthcoming paper \cite{GQS3}, $SM$ will be obtained as a co-adjoint orbit of
(a central extension of) the basic symmetry group above, and we shall perform the group quantization according to the algorithm
developed in \cite{GAQ} (see also \cite{Vallareport} and references there in).

For the sake of completeness, let us mention that the symmetry (\ref{simetria}), though being enough to achieve quantization, is not the full symmetry of
the system. In fact, it can be added with ``ordinary'' rotations generated by the Hamiltonian functions (Noether invariants):
\[ J_i\equiv\frac{1}{2}\eta_{ijk}\varepsilon^j\pi^k \]
along with the non-independent ones
\[ \kappa_i\equiv\rho\pi_i \]
closing now an Euclidean group $E(4)$. In fact, the combinations $J_i +\kappa_i$ and  $J_i -\kappa_i$ prove to be the Noether invariants
$\theta^R_i\equiv\vartheta_i$ and $\theta^L_i$ associated with the ``right'' and ``left'' SU(2) generators, leaving invariant the {\it chiral}
Lagrangian (\ref{Lagrangian}). This bigger (and non-minimal in the sense that it does not constitute the minimal generalization of the Heisenberg-Weyl group)
symmetry were pointed out in Ref.\cite{Isham} as a possible group to undertake the quantization of the $S^3$ particle
according to the Wigner-Mackey algorithm of induced representations of semi-direct product groups \cite{Wigner-Mackey}. There, although the realization of
this symmetry was not explicit, one would have also been lead to non-point transformations which do not leave the Lagrangian invariant, in contrast with
the original aim.

By making use of the inverse HJ transformation both symmetries and corresponding Noether invariants can be ``lifted'' to the EM.
In fact, the basic symmetries adopt the expressions:
\bea
\mathbb{X}_{(i)}&\equiv& X_{(i)}^1=\bar{X}_{(i)} =
X^k_{(i)}\frac{\!\!\partial}{\partial\epsilon^k} + \eta^k_{.ij} \theta^j \frac{\!\!\partial}{\partial\theta^k} \nn
\\
\mathbb{Y}_{(j)}&\equiv& Y_{(j)}^1= \frac{1}{\omega}
\left(
(\delta^n_j-\frac{1}{4}\epsilon_j\epsilon^n-\frac{1}{4\omega^4}X_{(k)j}\theta^kX_{(s)}^n\theta^s)
\hbox{sin}(\omega t)+
    \frac{1}{4\omega} X_{(l)}^n\theta^l X_{(r)j}(\vec{\varepsilon}\,)\theta^r t \right)
    \frac{\!\!\partial}{\partial\epsilon^n} \nn\\
     &+&
    \left(X_j^{(n)}\hbox{cos}(\omega t)
    +\frac{1}{\omega}
    (
     \frac{1}{2}\eta^n_{.jm}X_{(r)}^m\theta^r + \frac{1}{4}\delta_j^n \theta_k\epsilon^k
    )\hbox{sin}(\omega t)
      \right)\frac{\!\!\partial}{\partial\theta^n} \nn \\
\mathbb{Z}&\equiv& Z^1=-\frac{\varepsilon^k\mathbb{Y}_{(k)}}{\rho}\label{simetriaarriba}\,,
\eea
which are definitely not prolongation of transformations on $\mathbb{R}\times S^3$, except for  the generators of the group $SU(2)$.
In (\ref{simetriaarriba}) we have used for short the function $X_{(r)j}(\vec{\varepsilon}\,)=
\sqrt{1-\frac{\vec{\varepsilon}\,^2}{4}}\delta_{rj}+\frac{1}{2}\eta_{jrm}\varepsilon^m$
as well as the function $\rho$ defined above as depending on $\vec{\varepsilon}$.

The vector fields $\mathbb{X}_{(i)},\;\mathbb{Y}_{(j)}$, along with the inescapable companion $\mathbb{Z}$,
constitute the generators of the basic symmetry for the $S^3$ particle Sigma Model; here  $\mathbb{X}_{(i)}$ play the role of ``translations'',
whereas $\mathbb{Y}_{(j)}$ behave as ``boosts''. In fact, it should be noticed that if we would
have made the sphere radius $R$ in $S^3$ explicit, by redefining
$\vec{\epsilon}\rightarrow\frac{2}{R}\vec{\epsilon}$, the realization of
$\mathbb{X}_{(i)},\mathbb{Y}_{(j)}$ in (\ref{simetriaarriba}) goes to that of the (unextended) Heisenberg-Weyl algebra 
when $R\rightarrow\infty$, whereas the field $\mathbb{Z}$ is realized by $0$.


\section{Summary, outlook, final remarks}

In this paper he have pointed out the essential role played by the
Solution Manifold in the structure of a mechanical system
characterized by a Lagrangian and/or a Poincar\'e-Cartan form,
usually associated with it. In particular, the tangent/cotangent
space defined in terms of symmetries. To achieve this construction
we have needed the extension of ordinary (point) symmetries to
more general contact ones. In the midway we have learned the
symmetry sharing mechanism among physical systems bearing globally
diffeomorphic solution manifolds. But the main interest of these
considerations is really the establishment  of the basis for the
correct quantization in those cases (of non-linearity) where the
success of the Canonical Quantization algorithm is not guaranteed.
In fact, we should aim at finding a symmetry group of the physical
system among the co-adjoint orbit of which we can find our
Symplectic Solution Manifold. This way, the unitary, irreducible
representations of this group associated with the particular
co-adjoint orbit fulfils the correct quantization task. Also in
the middle of the way we have characterized, from the strict
classical framework, aspects of the perturbation scheme like the
Interaction Picture, versus the Schr\"odinger or Heisenberg ones which
would be associated with the classical description in EM or SM respectively.

It is worth mentioning, in relation to the symmetry quantization approach,
that the standard Canonical Quantization can be seen as a particular case of symmetry
quantization associated with the Heisenberg-Weyl group in those cases where this
symmetry is able to fully and globally characterize the Solution Manifold.
This is the case of free systems or not so free but sharing the
Heisenberg-Weyl symmetry; in that case we do not expect, however, a standard
realization of the symmetry generators and, then, basic operators.

It has been remarked in this paper that finding the  basic
symmetries of a given physical system is in general tantamount to
finding the explicit solutions of it. Nevertheless, in physical
systems characterized in its own essence by a certain symmetry, as
occurs indeed with fundamental interactions (in field theory), we
may figure out the actual structure of the solution manifold and
write the explicit form of the Hamiltonian (either classical or
quantum) and approach the time evolution with the Magnus series
perturbative scheme here sketched. Another perturbative scheme
would consist in trying to close, order by order in a
characteristic coupling constant $\lambda$, a  Lie algebra (at all
orders, infinite-dimensional , indeed) with the Hamiltonian and
finding the corresponding unitary and irreducible representations
at each order in $\lambda$. This new scheme is being developed at
present and will be communicated soon \cite{to appear}.


\section*{Acknowledgments}

Work partially supported by the  Spanish MICINN,
Junta de Andaluc\'\i a and Fundaci\'on S\'eneca,  under projects  FIS2011-29813-C02-01, FQM219-FQM4496 and 08814/PI/08, respectively.
Discussions with P. Horvathy are also acknowledged.

\end{document}